\documentclass[aps,prb,preprint,superscriptaddress]{revtex4-1}
\usepackage{amsmath}    % need for subequations
\usepackage{breqn}      % for multiple line equations
\usepackage{graphicx}   % need for figures
\usepackage{verbatim}   % useful for program listings
\usepackage{color}      % use if color is used in text
\usepackage{subfigure}  % use for side-by-side figures
\usepackage{hyperref}
\usepackage{epstopdf}

\newcommand{\ket}[1]{\ensuremath{\lvert #1 \rangle}}

\newcommand{\oa}{\ensuremath{\Omega_a}}
\newcommand{\ob}{\ensuremath{\Omega_b}}

\newcommand{\ga}{\ensuremath{\Gamma_a}}
\newcommand{\gb}{\ensuremath{\Gamma_b}}
\newcommand{\gp}{\ensuremath{\Gamma_p}}
\newcommand{\ave}[1]{\ensuremath{\langle #1 \rangle}}
\newcommand{\ci}[1]{\ensuremath{\langle\chi_{#1}\rangle}}
\newcommand{\da}{\ensuremath{\delta_a}}
\newcommand{\db}{\ensuremath{\delta_b}}

\begin{document}

\title{Sensing Coherent Phonons with Two-photon Interference}
\author{Ding Ding}
\affiliation{Department of Mechanical Engineering$,$ University of Colorado Boulder$,$ Boulder$,$ CO 80309$,$ USA}
 %\email{ding.ding-1@colorado.edu}
\affiliation{Singapore Institute of Manufacturing Technology$,$ 2 Fusionopolis Way$,$ Singapore 138634$,$ Singapore}
 \author{Xiaobo Yin}
\affiliation{Department of Mechanical Engineering$,$ University of Colorado Boulder$,$ Boulder$,$ CO 80309$,$ USA}
\author{Baowen Li}
 \email{Baowen.Li@colorado.edu}
\affiliation{Department of Mechanical Engineering$,$ University of Colorado Boulder$,$ Boulder$,$ CO 80309$,$ USA}


\date{\today}% It is always \today, today,

\begin{abstract}
Detecting coherent phonons pose different challenges compared to coherent photons due to the much stronger interaction between phonons and matter. This is especially true for high frequency heat carrying phonons, which are intrinsic lattice vibrations experiencing many decoherence events with the environment, and are thus generally assumed to be incoherent. Two photon interference techniques, especially coherent population trapping (CPT) and electromagnetically induced transparency (EIT), have led to extremely sensitive detection, spectroscopy and metrology. Here, we propose the use of two photon interference in a three level system to sense coherent phonons. Unlike prior works which have treated phonon coupling as damping, we account for coherent phonon coupling using a full quantum-mechanical treatment. We observe strong asymmetry in absorption spectrum in CPT and negative dispersion in EIT susceptibility in the presence of coherent phonon coupling which cannot be accounted for if only pure phonon damping is considered. Our proposal has application in sensing heat carrying coherent phonons effects and understanding coherent bosonic multi-pathway interference effects in three coupled oscillator systems. 

\end{abstract}

\maketitle

Phonons are packets of vibrational energy that shares many similarity with its bosonic cousin photons. Advances in nanofabrication has enabled many parallels between the development of photon and phonon control. Parallel developmets in passive control techniques include photonic \cite{Joannopoulos1997} versus phonoic crystals \cite{III2009}, optical \cite{Cai2010} versus acoustic metamaterials \cite{Ma2016} etc.  Development in active manipulation of electromagnetic waves through light-matter interaction have led to creation of nanoscale optical emitters \cite{Willander2014} and gates \cite{Chen2013} and similar progress have been made in controlling phonons using their interaction with matter especially in the realms of optomechanics \cite{Aspelmeyer2014} and phononic devices\cite{Li2012,Han2015}. Phonons span a vast frequency range and while techniques to control and sense lower frequency coherent phonons have been well-developed \cite{Ikezawa2001,Lanzillotti-Kimura2007,Vahala2009,Grimsley2011,Hong2012,Tian2014,Wang2014,Yoshino2015,Volz2016,Shinokita2016}, heat carrying coherent terahertz acoustic phonons have been harder to measure directly due to the small wavelength and numerous scattering mechanism at these small wavelengths \cite{Chen2005}. In the past, THz crystal phonons have been generated and detected in low temperature experiments with defect doped crystals \cite{Renk1971,Renk1979,Bron1980,wybourne_phonon_1988}, with experimental evidence of coherent phonon generation \cite{Bron1978,Hu1980,Fokker1997}. At the same time, interpretation of non-equilibrium phonon transport, with the advancement of nanoscale electrical heating and ultrafast optical pump-probe techniques, have allowed us to infer phonon coherence from broadband thermal conductivity measurements \cite{highland_ballistic-phonon_2007,Luckyanova2012,Ravichandran2014,Latour2014,Alaie2015}. There have been also interest of using defect-based techniques as a thermal probe using perturbation to energy levels due to changes in temperature \cite{laraoui_imaging_2015}. Furthermore, surface deflection techniques with ultrafast optics have also been used to generate phonons close to THz frequencies in materials \cite{Kent2002,Kent2006,Cuffe2013,Maznev2013}. Defect-based techniques are attractive compared to both thermal conductivity measurement and deflection techniques due to its ability to directly access atomic length scales where THz phonon wavelength resides. Also, the energy levels in the excited state electron manifold of these defects can match the phonon energy precisely \cite{Sabisky1968,Renk1971,Eisfeld1979}, resulting in a narrow band phonon detector. 

In light of the success of defect-based optical absorption techniques in coupling directly to high frequency phonons, we propose the use of two photon interference to measure the coherence properties of these phonons. Two photon interference techniques, with the most famous being coherent population trapping (CPT) \cite{Arimondo1996} and electromagnetically induced transparency (EIT) \cite{Fleischhauer2005}, have been widely adopted in spectroscopy and metrology in atomic \cite{Janik1985,duong_measurement_1974} and defect-based systems \cite{Zhao1997,Ham1998,Hemmer2000,Acosta2013,rogers_all-optical_2014}. However, CPT and EIT usually excludes the possibility of a ground state coupling \cite{Whitley1976} or merely treating the ground state coupling as thermal bath \cite{Acosta2013}. 

In this paper, we propose the possibility of using the presence of coherent coupling of two ground states in a $\Lambda$ system by THz acoustic phonons of the host material as a coherent phonon sensor. We show two experimentally observable effects, namely an asymmetric excited state population lineshape in CPT and an anomalous dispersion profiles in EIT measurements, which only occurs in the presence of coherent phonon coupling to a lattice phonon mode. Our proposal has the potential for direct implementation in defect-based phonon detection experiments mentioned earlier \cite{Sabisky1968,Renk1971,Eisfeld1979} and extends traditional two couple oscillator models in two photon interference to a three-coupled-oscillator models \cite{cosmelli_asymmetries_1993,Fink2000}. Our result will also be applicable for three-way coupled system such as microwave driven quantum-beat lasers \cite{Scully1985,Swain1988}, designed opto/electro-mechanical schemes \cite{Hatanaka2013,Sollner2016} or phonon-based quantum memories \cite{Hong2012,England2013,Albrecht2013}.

\begin{figure*}[htpb]
\centering
\includegraphics[width=\textwidth]{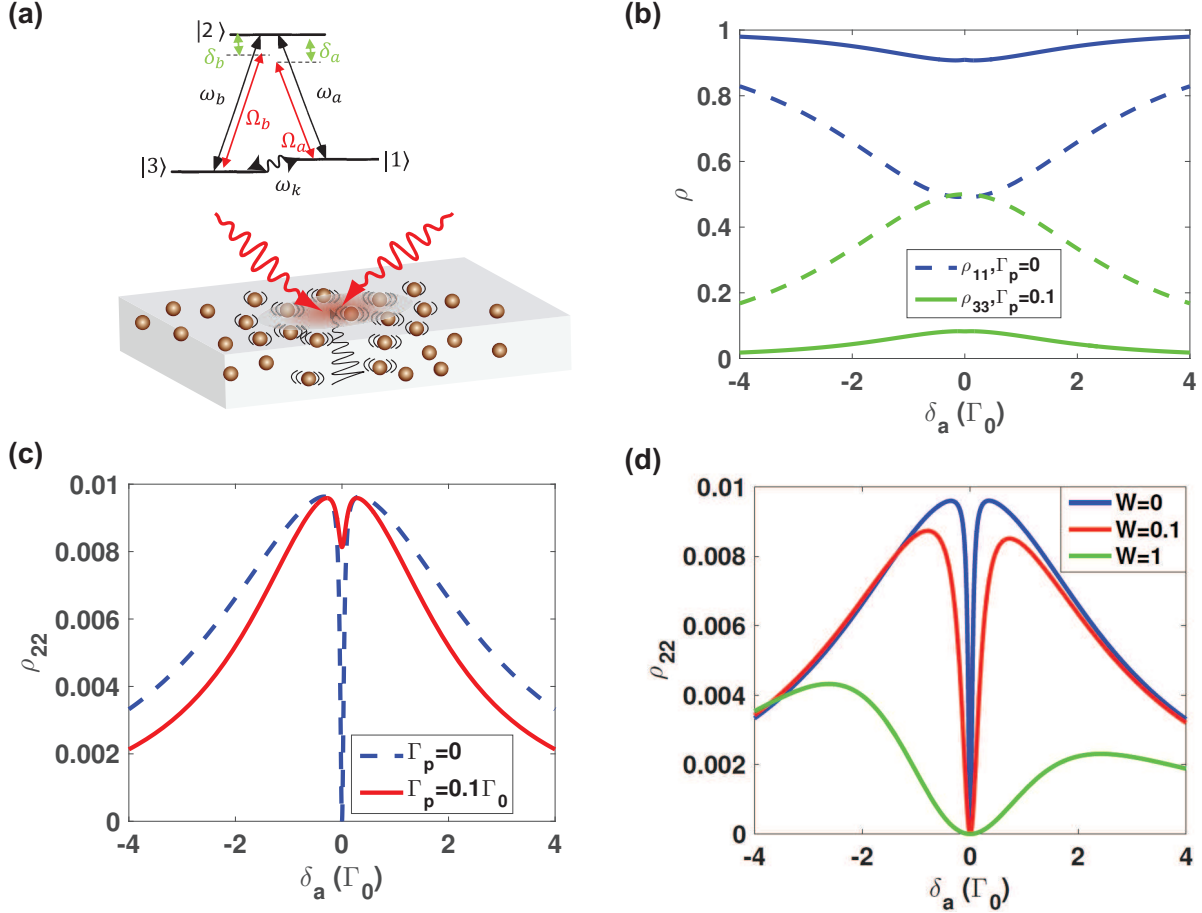}
\caption{\label{fig:schematic} (a) Schematic of two photon interference in a defect-based crystalline system. The emitters are identical and have a $\Lambda$ type energy level system. Levels $\ket{1}$ and $\ket{3}$ are part of the ground state manifold and the excited state $\ket{2}$ have a frequency difference of $\omega_a$ and $\omega_b$ respectively. The optical fields driving the $\ket{2}-\ket{1}$ transition and that driving $\ket{3}-\ket{2}$ transition have frequency $\oa$ and $\ob$ with detunings $\da$ and $\db$ respectively. The phonon field create vibrations to the defect emitters and results in a resonant coupling between the ground states $\ket{3}$ and $\ket{1}$. (b) Populations in states $\ket{1}$ (blue lines) and $\ket{3}$ (green lines) as a function of the detuning $\da$. The dashed line represents the case with no phonon damping. The solid line represents the case with phonon damping $\gp=0.1\Gamma_0$. The optical coupling strength for both $\ket{2}-\ket{1}$ and $\ket{3}-\ket{2}$ transition are $G=0.2\Gamma_0$, where $\ga=\gb=\Gamma_0$ for the optical damping. (c) Population in level 2 with and without damping represented by solid and dashed line (Eq. \ref{eq:r22_w0}) respectively. There is a sharp dip to zero population at $\da=0$ which is due to two-photon interference. (d) Population versus detuning $\da$ in excited state $\ket{2}$ for different phonon coupling strength $W$ and no phonon damping.}
\end{figure*}

In the schematic of our proposal in Fig. \ref{fig:schematic}(a), a two-photon interference is created in a localized region of a medium that carries an ensemble of identical emitters with electronic energy level resembling a typical $\Lambda$ system used in CPT or EIT. The optical fields driving the $\ket{2}-\ket{1}$ and $\ket{3}-\ket{2}$ transitions have detunings $\delta_a$ and $\delta_b$ with respect to the electronic energy levels of the emitters. The total Hamiltonian of the system can be written as
\begin{widetext}
\begin{subequations}
\begin{align}
H=&H_A+H_F+H_I \label{eq:total_ham}\\
H_A \ket{m}=&E_m \ket{m} \label{eq:H_A} \\
H_F=&\hbar \sum_{\lambda} \omega_\lambda c_\lambda^{\dagger} c_\lambda + \hbar \sum_{k} \omega_k b_k^{\dagger} b_k \label{eq:H_F} \\
H_I=&\hbar \sum_\lambda \left( g_a^{\lambda} \sigma_{21} c_\lambda + g_b^{\lambda} \sigma_{23} c_\lambda + c.c. \right) +\hbar \sum_k \left( \zeta_k \sigma_{31} (b_k+b^{\dagger}_k)+c.c.\right) \label{eq:H_I}
\end{align}
\end{subequations}
\end{widetext}
where the electronic part satisfies the eigenvalue Eq. \ref{eq:H_A} of electronic eigenstate $\ket{m}$, the field part (Eq. \ref{eq:H_F}) is the usual expression that now comprises the sum of the photon modes indexed as $\lambda$ with raising and lowering operators $c^{\dagger}_{\lambda},c_\lambda$ and the phonon modes indexed as $k$ with raising and lowering operators $b_k^{\dagger},b_k$. The interaction Hamiltonian in Eq. \ref{eq:H_I} has two parts, the first part being the original two photon interference Hamiltonian which realizes effects of CPT and EIT, and the other portion responsible for phonon interaction. 

Using the procedure outlined in Supplementary Information (SI) similar to the method by Whitney and Stroud \cite{Whitley1976}, one arrives at the set of Eqs. S13 which specifies the equations of motion for elements of the density matrix. Note that in Eqs. S13, we are able to obtain spontaneous $\Gamma$ (Eqs. S11,S12) and stimulated rates $G_i,W$ Eq. S15 directly from the equations of motion Eq. S4 without having to add damping terms unlike semi-classical approaches and this is the merit of the approach by Whitley and Stroud \cite{Whitley1976}. The spontaneous damping terms are defined as sum over all mode contributions in both optical (Eq. S11) and phonon cases (Eq. S12) while the coherent optical coupling terms $G_{a,b}$ are defined for coupling to a specific mode $\alpha,\beta$ (Eqs. S15a and S15b) and $W$ for the specific phonon mode $\gamma$. A very important feature of our system is that we have now included the possibility for a coherent phonon coupling of strength $W$ that couples to the $\ket{3}-\ket{1}$ transition instead of a pure phonon damping term, and examining this feature will be the main theme of subsequent results and discussions. We would especially like to bring your attention to the definition of $W$ in Eq. S15c where ensemble average of the phonon annihilation operator will only yield a non-zero value if the detected phonons are coherent \cite{Whitley1976}. This is because an incoherent or thermal ensemble will yield a zero ensemble average \cite{scully_quantum_1997}. Thus, our proposed technique  offer a rigorous detection of phonons rather than indirect evidence using thermal conductivity measurements.

The diagonal terms $\rho_{11}$, $\rho_{22}$ and $\rho_{33}$ are the population of each energy level. We first solve for the steady state solution to Eq. S13 which allows us to obtain $\rho_{11}$, $\rho_{22}$ and $\rho_{33}$ in the long-time limit. We first consider CPT where the optical field for $\ket{2}-\ket{1}$ transition is tunable while transition $\ket{3}-\ket{2}$ is fixed, and that both fields are of equal strength $G_a=G_b=G$. Under the condition of no phonon damping $\Gamma_p=0$, unity optical damping $\ga=\gb=\Gamma_0$ and coupling $W=0$, we can obtain the expression of $\rho_{11}$, $\rho_{22}$ and $\rho_{33}$ as 
%,
\begin{subequations}
\begin{align}
\rho_{11}=&\frac{1}{2}\left(1+\frac{\da(\da^3-4\da G^2)}{\da^4+8G^4+2\da^2(4+G^2)}\right) \label{eq:r11_w0}\\
\rho_{22}=&\frac{2\da^2 G}{\da^4+8G^4+2\da^2(4+G^2)} \label{eq:r22_w0}\\
\rho_{33}=&\frac{1}{2}\left(1-\frac{\da^4}{\da^4+8G^4+2\da^2(4+G^2)} \right) \label{eq:r33_w0}
\end{align}
\end{subequations}

The dashed lines in Fig. \ref{fig:schematic} (b) plots the population of level $\ket{1}$ (Eq. \ref{eq:r11_w0}) and level $\ket{3}$ (Eq. \ref{eq:r33_w0}) which are in the ground state manifold. There is a broad resonance that peaks at zero detuning where almost half of the population is in each of the ground state. The excited state population of level $\ket{2}$ in Eq. \ref{eq:r22_w0} in Fig. \ref{fig:schematic} (c) is small for all detuning, where the dashed line also shows a broad resonance peak. However, there exist a sudden dip at $\delta_a=0$ to zero population, a feature of complete two photon resonance in CPT \cite{Gray1978,Janik1985,Arimondo1996}. Now, let us add some phonon damping $\Gamma_p=0.1\Gamma_0$ but assume no phonon coupling i.e. $W=0$. The solid lines in Fig. \ref{fig:schematic}(b) shows the population of level $\ket{1}$ and level $\ket{3}$ again where adding phonon damping reduces the population transfer between $\ket{1}$ and $\ket{3}$ at $\da=0$, leaving only ~10\% of population in level $\ket{3}$ on resonance. Fig. \ref{fig:schematic}(c) show that two photon interference effect in the excited state $\ket{2}$ with (solid line) phonon damping is reduced on resonance. This is physically expected as $\gp$ is a source of decoherence which reduces the ideal result in CPT or EIT.

Next, we introduce coherent phonon coupling $W$ and ignore phonon damping $\gp$ for the excited state level $\ket{2}$ given by Eq. S19. Figure \ref{fig:schematic}(d) shows the excited state population $\rho_{22}$ for different values of $W$. When $W$ is small, there is no noticeable change between the lineshape versus that in Fig. \ref{fig:schematic}(c) where $W=0$. However, as we increase $W$, then asymmetry starts to emerge. First, the position of the peak for positive and negative detuning $\da$ are shifted further apart as $W$ increases. Second, the difference between the maximum peak amplitude on the positive and negative detunings becomes greater as $W$ increases. Third, the original two photon resonance dip at $\da=0$ still remains at the same location and goes all the way to zero population for all values of $W$, implying the preservation of a dark state that is characteristic of CPT \cite{Arimondo1996}. These observations are very interesting so let us understand them one at a time. 

\begin{figure*}[htpb]
\centering
%\subfigure[]{
%\includegraphics[width=0.38\textwidth]{colormap_low_res.eps}}
%\subfigure[]{
%\includegraphics[width=0.38\textwidth]{neg_root_vary_w.eps}}
%\subfigure[]{
%\includegraphics[width=0.38\textwidth]{pos_rt_vary_w.eps}}
%\subfigure[]{
%\includegraphics[width=0.38\textwidth]{diff_rho22.eps}}
\includegraphics[width=\textwidth]{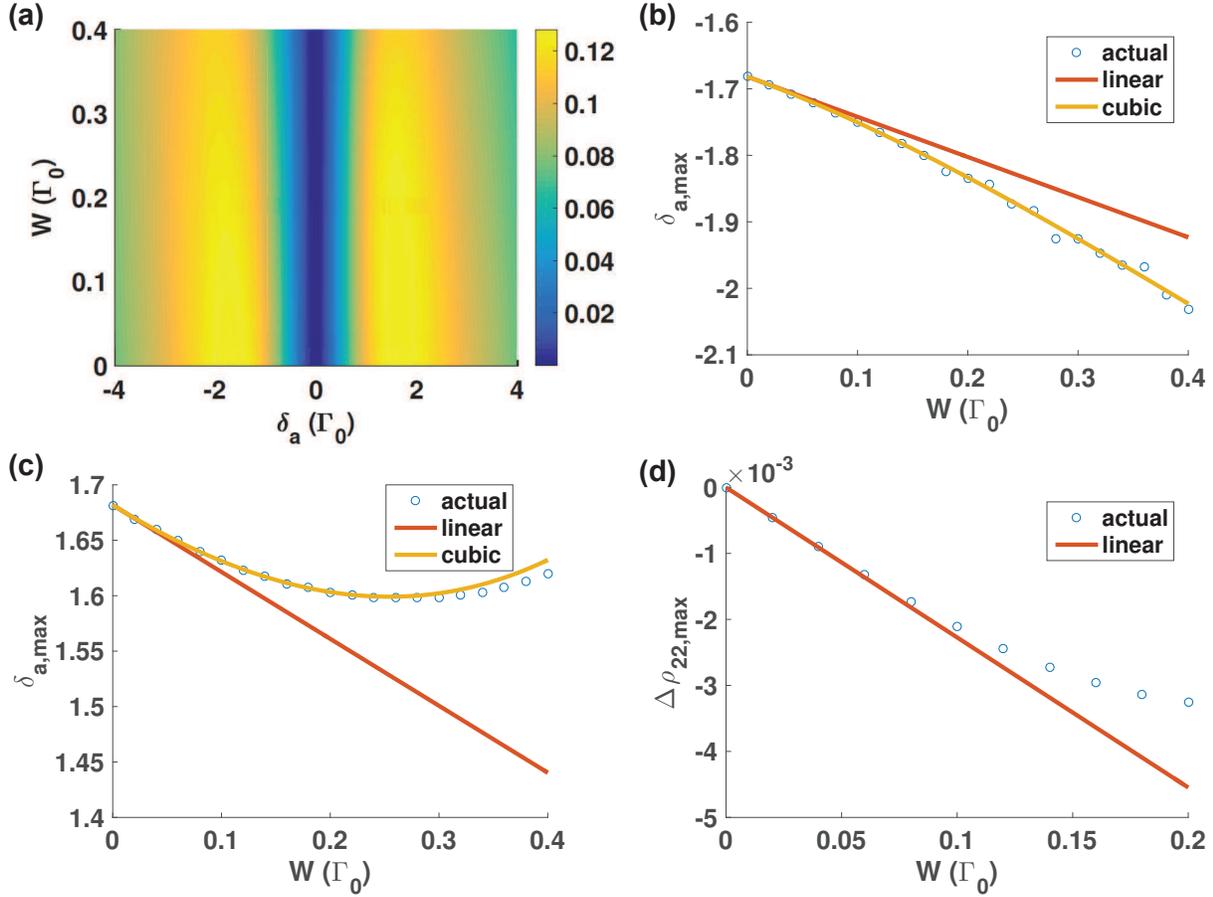}
\caption{\label{fig:position} (a) Two-dimensional plot of Population in level $\ket{2}$ as a function of detuning $\da$ and phonon coupling $W$ for $G=\Gamma_0, \ga=\gb=\Gamma_0$. The yellow region on both positive and negative detuning are the maximum positions while the dark blue region at $\da=0$ indicates the resonance dip just like in Figs. \ref{fig:schematic}(c,d). The variation of the positions of the maximum detuning $\delta_{a,max}$ are plotted as blue circles in (b) and (c) for negative and positive detuning respectively. The linear relation between maximum position $\delta_{a,max}$ for small $W$ can be related to the linear term in Eq. S20 while higher order terms account for the variation in maximum position and phonon coupling $W$. (d) Difference between negative and positive peak height as a function of phonon coupling $W$. The linear term in Eq. S20 accounts for the trend for small phonon coupling $W\lesssim 0.1\Gamma_0$.} 
\end{figure*}

To understand the first and second observation, we map the variation of the excited state population $\rho_{22}$ as a function of $W$ and detuning $\da$ for a larger value of $G=\Gamma_0$ in Fig. \ref{fig:position}(a), with no phonon damping ($\Gamma_p=0$) and unity photon damping ($\Gamma_a=\Gamma_b=\Gamma_0$) using Eq. S19. A larger value of $G$ compared to Fig. \ref{fig:schematic} allows us explore a wider range of values for $W$ in the range of $W\ll G$ to $W\sim G$. As evident from Fig. \ref{fig:position}(a), the two yellow regions indicating the negative and positive detuning maxima vary with $W$. The blue circle in Figs. \ref{fig:position}(b) and (c) show that the negative detuning maxima and positive detuning maxima in Fig. \ref{fig:position}(a) as a function of increasing $W$, respectively. The trend in Figs. \ref{fig:position}(b) and (c) can be explained analytically by looking at the solution of the turning points for the steady-state solution of $\rho_{22}$ (Eq. S19). There are three turning points, where one is at $\da=0$ which is the CPT resonance in Figs. \ref{fig:schematic}(c,d). The other two turning points can be described by Taylor expansion of Eq. S19 as a function of $W$, resulting in Eq. S20 where $\da=\pm 2^{3/4} G$ is the zeroth order solution which are symmetric about $\da=0$ (as in Figs. \ref{fig:schematic}(c,d)). For small $W$, the linear term $- \frac{1}{2} \left( \frac{1}{2}+\frac{1}{\sqrt{2}} \right) W$ in Eq. S20 dominates and Figs. \ref{fig:schematic}(c,d) show both linearly decreasing trend for $W\lesssim 0.1\Gamma_0$ (shown in red solid line in Figs. \ref{fig:position}(b,c)). However, when $W$ is increased further, then the higher order terms in Eq. S20 starts to dominate, increasing the positive maximum value and decreasing the negative maximum, consistent with the observation of the shift in detuning as $W$ increases in Fig. \ref{fig:position}(b,c). 

Next, we examine how phonon coupling $W$ creates asymmetry in the peak heights in Fig. \ref{fig:position}(b). We substitute the linear term in Eq. S20 into the steady state solution for $\rho_{22}$ (Eq. S19) to obtain difference between the positive and negative detuning as

\begin{equation}
\Delta\rho_{22,max}\approx\frac{2^{5/4}(\sqrt{2}-2)G^3 W}{16 + 8 G^2 + 16 \sqrt{2} G^2 + 9 G^4 + 4 \sqrt{2} G^4} \label{eq:rho22max}
\end{equation}
. Equation \ref{eq:rho22max} is plotted as a function of $W$ in Fig. \ref{fig:position}(d) to show that the linear regime agrees well with the actual data from Fig. \ref{fig:position}(a) for small values of $W$. Experimentally, this linearity allows direct retrieval of the value of phonon coupling $W$ from experimental measurements of excited state population $\rho_{22}$ if optical fields couplings are much stronger than phonon coupling $G\gg W$. 

The third observation is the preservation of the resonance dip to zero occupation in Fig. \ref{fig:position} for all $W$, indicating that the dark state is preserved just like in the CPT case in Fig. \ref{fig:schematic}(c). The dressed state picture allows us to identify the eigenstates by diagonalizing the interaction Hamiltonian in Eq. \ref{eq:H_I} with $G_a=G_b=G$ on resonance (i.e. $\da=\db=0$) \cite{Fleischhauer2005} in matrix form as 

\begin{equation}
H_I=
\begin{bmatrix}
0 & W & G \\
W & 0 & G \\
G & G & 0
\end{bmatrix}
\label{eq:H_I_mat}
\end{equation}

where the dressed states can be obtained by taking the eigenvector and eigenvalues of Eq. \ref{eq:H_I_mat}. In the absence of phonon coupling where $W=0$, we obtain the familiar dressed state result of a CPT system \cite{Fleischhauer2005} where the eigenvalues are $(0,\pm\sqrt{2}G)$ and the eigenvectors are

\begin{subequations}
\begin{align}
\ket{a_0}=&\ket{3}-\ket{1} \label{eq:a_0}\\
\ket{a_{\pm}}=&\ket{1}+\ket{3}\pm\sqrt{2}\ket{2} \label{eq:a_pm}
\end{align} \label{eq:dressed_w0}
\end{subequations}

. Equation \ref{eq:a_0} is the dark state as it does not contain any excited state $\ket{2}$. Physically, this means that the ground states are mixed with no population in the excited state when the system is in a dark state. 

When $W$ is non-zero, the eigenvalues are modified to $(-W,1/2(W\pm\sqrt{8G^2+W^2}))$ and the eigenvectors become 

\begin{subequations}
\begin{align}
\ket{a_0}=&\ket{3}-\ket{1} \label{eq:a_0_w}\\
\ket{a_{\pm}}=&\ket{1}+\ket{3}\pm\frac{\sqrt{8G^2+W^2}\pm W}{2G}\ket{2} \label{eq:a_pm_w}
\end{align} \label{eq:dressed_w}
\end{subequations}

. Equation \ref{eq:dressed_w} shows that the dark state $\ket{a_0}$ is preserved even when $W$ is non-zero. This is consistent with the observation of the preservation of the dip on resonance despite the presence of phonon coupling in Figs. \ref{fig:schematic} and \ref{fig:position}. However, the eigenvalue of the dark state is now $-W$ instead of $0$, implying a different time evolution of the eigenstates compared to the case in Eq. \ref{eq:dressed_w0} where $W=0$.

\begin{figure*}[htpb]
\centering
\includegraphics[width=\textwidth]{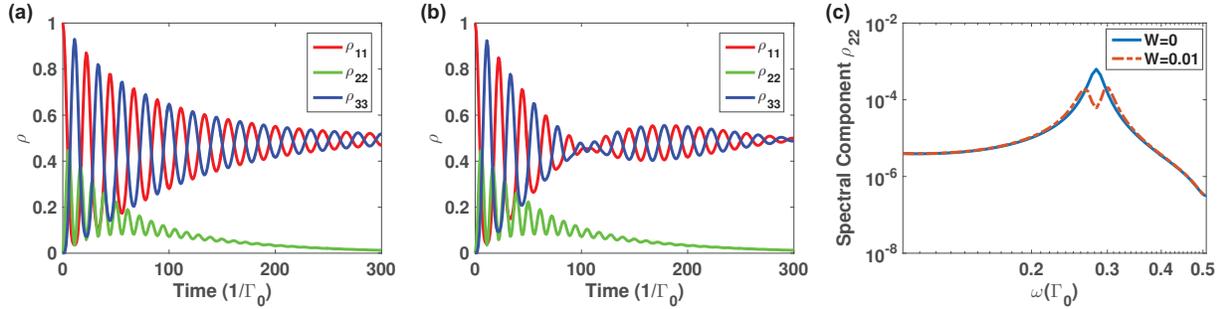}
\caption{\label{fig:time} (a) Time evolution of population in levels $\ket{1}$, $\ket{2}$ and $\ket{3}$ with no phonon coupling and no phonon damping. Optical damping $\ga=\gb=0.01\Gamma_0$ is set smaller than in previous figures so as to showcase more oscillatory features of the evolution. The optical coupling is the same as the CPT case in Fig. \ref{fig:schematic} where $G=0.2\Gamma_0$. All population tend to the steady state values predicted in Fig. \ref{fig:schematic} for long evolution times. (b) Time evolution of population in levels $\ket{1}$, $\ket{2}$ and $\ket{3}$ with phonon coupling $W=0.01\Gamma_0$ and no phonon damping. The time oscillations are now modulated at a slower frequency especially for level $\ket{1}$ and level $\ket{3}$ shown in blue and yellow respectively. (c) Fourier transform of population in level $\ket{1}$ for the case of no phonon coupling (a) and with phonon coupling (b). The blue solid line shows that when there is no phonon coupling, there is a peak oscillation at $\sqrt{2}G$. However, when phonon coupling is present, the red dashed line shows that the fundamental oscillation frequency is now split into two frequencies due to the presence of phonon coupling.  }
\end{figure*}

Now, we examine time dynamics of the electronic populations in levels $\ket{1}$,$\ket{2}$ and $\ket{3}$ on resonance (i.e. $\da=0$) in the presence of phonon coupling. The details on how to obtain the population time dynamics is given in the SI. Under light optical damping $\ga=\gb=0.01\Gamma_0$ and zero phonon coupling and damping $W,\Gamma_p=0$, many optical oscillation persist as demonstrated in Fig. \ref{fig:time}(a). The populations $\rho_{11}(t),\rho_{33}(t)$ tend to 0.5 which is the steady state value in Fig. \ref{fig:schematic}, likewise for $\rho_{22}(t)$ in after $t=300/\Gamma_0$. The Fourier transform of $\rho_{11}(t)$ (blue solid line in in Fig. \ref{fig:time}(c)) shows a peak at $\sim0.28\Gamma_0$. The peak matches almost the value of $\sqrt{2}G$ where $G=0.2\Gamma_0$ as expected in CPT \cite{Arimondo1996} and from Eq. \ref{eq:dressed_w0} \cite{Fleischhauer2005}. However, with non zero phonon term $W=0.01 G_a$, $\rho_{11}(t)$ and $\rho_{33}$ both have a slower modulation on top of the faster optical oscillation as shown by the blue and yellow lines of population in levels $\ket{1}$ and $\ket{3}$ in Fig. \ref{fig:time}(b). If we take the Fourier transform of $\rho_{11}(t)$ again, we obtain the red dashed spectrum in Fig. \ref{fig:time}(c) where the first peak now shows a splitting of frequency with respect to the undisturbed case. The splitting into two frequencies at $\omega_+\sim0.27\Gamma_0$ and $\omega_+\sim0.29\Gamma_0$ resembles the splitting in eigenvalues $1/2(W\pm\sqrt{8G^2+W^2})$ of eigenvectors in Eq. \ref{eq:dressed_w}. Physically, phonon coupling $W$ results in non-degenerate eigenvalue magnitudes such that $\ket{a_+}$ and $\ket{a_-}$ oscillate at different eigenfrequencies. This in turn modulates population $\rho_{11}(t)$ and $\rho_{33}(t)$, causing a splitting of the frequency compared to the case where phonon coupling $W=0$.  
 
\begin{figure*}[htpb]
\centering
\includegraphics[width=\textwidth]{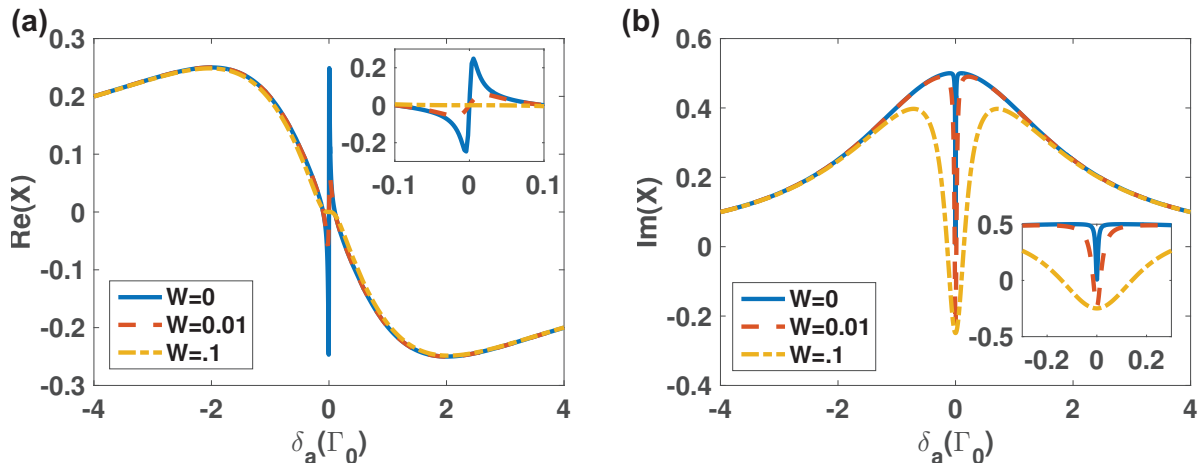}
\caption{\label{fig:eit} (a) Plot of real part of linear susceptibility $X$ as a function of detuning $\da$ for EIT where $\ga=\gb=\Gamma_0$ and $G_a=0.1\Gamma_0$ for different values of $W$. When $W=0$, the lineshape resembles a typical EIT lineshape \cite{Fleischhauer2005} and increasing $W$ decreases the sharpness of the turning points similar to the effect of increased damping. (b) Plot of real part of linear susceptibility $X$ as a function of detuning $\da$ with the same conditions as (a). Increasing $W$ leads to a negative value of the imaginary susceptibility which is anomalous compared to the case where $W=0$ where the imaginary part just goes to zero.   }
\end{figure*} 
 
Having looked at the CPT case, one wonders if we can use EIT technique to sense coherent phonons. In EIT, the condition for the optical fields becomes $G_a \ll G_b$, where the $\ket{2}-\ket{1}$ optical field is a now a weak probe with detuning $\da$ compared to a strong resonant driving field for the $\ket{3}-\ket{1}$ transition. The quantity of interest in EIT is the susceptibility of the medium \cite{Fleischhauer2005} under the incidence of the probe beam which is related to the off-diagonal steady state solution to the density matrix term $\ave{\ci{21}}$ in Eq. S14. Under the condition of no phonon field and damping $W=0, \gp=0$, we can obtain the linear susceptibility $X$ by Taylor expansion of the steady state solution for Eq. S17 for $\ci{21}$ for small $G_a$ to obtain 
\begin{equation}
X=\frac{\da}{G_b^2-i\ga\da-i\gb\da-\da^2} \label{eq:Xi}
\end{equation}
. Figure \ref{fig:eit} (a,b) plots the real and imaginary susceptibility for different values of $W$. The shape of the real and imaginary susceptibility for $W=0$ in Eq. \ref{eq:Xi} are typical EIT susceptibility \cite{Fleischhauer2005} showing a sharp inflection at zero detuning $\da=0$ for the real part and a sharp dip for the imaginary part. The dip to zero for the imaginary part (blue solid line in Fig. \ref{fig:eit}(b)) physically indicates zero absorption where the transparency window in EIT refers to. 

When we have phonon coupling $W>0$, we see changes in dispersion in Fig. \ref{fig:eit}(a,b). The change in the real part in Fig. \ref{fig:eit}(a) follows a decrease in the sharpness of the inflection which can also be due to effects of damping. However the negative anomalous imaginary part on resonance in Fig. \ref{fig:eit}(a) cannot be caused by damping. Damping will only reduce the size of the dip similar to the result of excited state population $\ket{2}$ in Fig. \ref{fig:schematic}(c). Thus, the presence of anomalous imaginary susceptibility at resonance is another good measure for the strength of phonon coupling $W$. Physically, negative anomalous imaginary susceptibility should indicate gain rather than loss, which means that we not only have transparency, but possibly amplification. The details of this possibility will be discussed in a future study.

Experimentally, this scheme offers a rigorous way to detect coherent phonons in the THz frequency range which is responsible for heat condition.  As mentioned earlier, these defect-based detection techniques have the characteristic of being narrow band and yet tunable \cite{Sabisky1968,Eisfeld1979} and has been employed successfully in understanding many aspects of phonon transport in crystals \cite{Bron1980} and interfaces \cite{Bron1977}. These crystals can be interfaced with other materials phonon detectors \cite{kaplyanskii_phonon_1984}, making our proposed method directly applicable to detecting coherent phonons in thermal transport. 

To experimentally realize our proposal, three challenges need to be addressed. Firstly, CPT or EIT have yet been experimentally demonstrated with THz energy separation between the ground state manifold to our knowledge. However, we believe that with the advent of frequency combs, locking two laser in the THz range is certainly possible \cite{consolino_phase-locking_2012} and we may soon see such an experiment being performed. Secondly, phase fluctuation in any of the optical or phonon fields will affect the quality of the photon-phonon interference. Experimental demonstrations of CPT and EIT typically use the same laser source to generate two frequencies \cite{Arimondo1996,Fleischhauer2005}, leading to the same phase fluctuations in both optical fields. Dalton and Knight \cite{Dalton1982} specifically addressed this issue for two photon interference where $\Lambda$ will be spared of any decoherence but not in a ladder system. Here, our two-photon-phonon interference is a composite of $\Lambda$ and ladder systems and the net effect will be a reduced interference effect. Lastly, due to phase fluctuation, the coherent phonon field must carry the same phase fluctuation as the optical field, so we must generate the phonons in a coherent manner with the same laser field for the $\ket{2}-\ket{1}$ and $\ket{3}-\ket{1}$ transitions. This is possible with the advent of coherent phonon sources in defect-based systems \cite{Bron1978,Hu1980,Fokker1997}, material systems \citep{Kent2002,Kent2006,Cuffe2013,Maznev2013} and nanofabricated systems \cite{Lanzillotti-Kimura2007,Grimsley2011,Hong2012,Tian2014,Wang2014,Yoshino2015,Volz2016,Shinokita2016}.

Our work differs from the field optomechanics and non-linear coherent phonon control \cite{kozak_coherent_2013}. Optomechanics primarily relies on coupling a mechanical mode to a designed optical cavity for coherent phonon control. It is remarkable that quantum coherence of phonons has been predicted \cite{hu_quantum_1996,hu_phonon_1999,hu_quantum_2015} and observed \cite{safavi-naeini_electromagnetically_2011,Aspelmeyer2014} in this field. Here, we are proposing a detection scheme with optical defects which couples to intrinsic crystal lattice phonon modes in materials. Also, we only restrict our discussion here to coherent and thermal state although it is possible to consider other quantum states such as Fock states and squeezed states \cite{hu_quantum_1996,hu_phonon_1999,hu_quantum_2015} . For the field of non-linear coherent phonon generation, an optical field directly couples to optical phonons \cite{kozak_coherent_2013} or zone-center acoustic phonons \cite{olsson_temperature_2015} and as a result of the phase matching, always results in coherent phonons being observed. Our work actually detects high frequency acoustic phonons which are not capable of direct coupling to light through phase matching. Furthermore, our technique can detect both coherent and incoherent phonons through their ensemble distribution and no phase matching is required. Recent work of share some similarity to ours include phonon mediated gate operations using defects in nitrogen vacancy centers \cite{Albrecht2013} and characterizing phonon coherence in thermal transport using correlation functions \cite{Latour2014}. It is thus evident that characterizing high frequency coherent acoustic phonons in materials using quantum mechanical description are only starting to be explored. 

Lastly, we would like to mention the relevance of our work not limited to phonon sensing, but also to three-way interference problems \cite{Hatanaka2013,Sollner2016} and coupled oscillator systems \cite{cosmelli_asymmetries_1993,Fink2000}. Our theory is not limited to just phonon coupling of the ground state manifold but any bosonic field. Thus, the predicted asymmetry in the excited state population, modulation in population time dynamics and the anomalous EIT dispersion will also be observable in any of the above systems, paving way to understanding and engineering multiple interference pathways in more complex multilevel systems.

In conclusion, we have proposed a coherent phonon sensing scheme that utilized the existing two photon interference techniques to rigorously test the presence of coherent phonons. Modifications to steady state population lineshape, modulation in ground state time dynamics, and anomalous EIT signal with negative imaginary susceptibility all provided a wealth of indicators for which coherent phonons can be sensed experimentally. Moreover, our scheme can be applicable to understanding other multi-interference phoenomona. The main advantages of our scheme is the ability for atomic scale emitters to sense small-wavelength terahertz coherent phonons in materials accurately and precisely, and that two-photon inference technique allows for a direct, sensitive and rigorous conclusion to the presence of coherent phonons in materials.

%
%\bibliography{ref_v2_bib}

\end{document}

\def\theequation{S\arabic{equation}}

\title{Supplementary Information}
\author{Ding Ding}
\author{Xiaobo Yin}
\author{Baowen Li}
 \email{Baowen.Li@colorado.edu}
%\affiliation{Department of Mechanical Engineering, University of Colorado Boulder, Boulder, CO 80309, USA}


\date{\today}% It is always \today, today,

\begin{abstract}
We provide the derivation to the manuscript.
\end{abstract}

\maketitle

\section{Derivation for system with phonon coupling}

This derivation follows closely the work of Whitley and Stround [49]. Consider a $\Lambda$ system described in the main text where the frequency difference between the $\ket{2}-\ket{1}$ transition is expressed as $\Omega_a=(E_2-E_1)/\hbar$ and the $\ket{2}-\ket{3}$ transition as $\Omega_b=(E_2-E_3)/\hbar$. The photon field driving the $\ket{2}-\ket{1}$ transition has a detuning $\delta_a=\Omega_a-\omega_a$ and the field driving the $\ket{2}-\ket{3}$ transition has a detuning of $\delta_b=\Omega_b-\omega_b$. The total Hamiltonian of the system can be written as
Eqs. 1a-d where the atomic part satisfies the eigenvalue Eq. 1b, the field part (Eq. 1c) is the usual expression that now comprises the sum of the photon modes  indexed as $\lambda$ with raising and lowering operators $c^{\dagger}_{\lambda},c_\lambda$ and the phonon modes indexed as $k$ with raising and lowering operators $b_k^{\dagger},b_k$. The interaction Hamiltonian in Eq. 1d has two parts, the first part being the original two photon interference Hamiltonian which realizes effects of CPT and EIT, and the other portion responsible for phonon coupling here. Notice that we did not use the rotating wave approximation for the phonon part. The coupling coefficient $g_a^{\lambda}$ and $g_b^{\lambda}$ are stands for interaction of the photon dipole interaction for the $\ket{2}-\ket{1}$ and $\ket{2}-\ket{3}$ transition in Fig. 1(a) respectively and the coupling coefficient $\zeta_k$ stands for electron phonon interaction. The magnitude of the coupling constants are given by

\begin{align}
g_a^{\lambda}=&-i e \mathbf{\hat{\epsilon_\lambda}\cdot d_a} \sqrt\frac{2\pi\omega_\lambda}{\epsilon_0\hbar V_l} \label{eq:g_a}\\
g_b^{\lambda}=&-i e \mathbf{\hat{\epsilon_\lambda}\cdot d_b} \sqrt\frac{2\pi\omega_\lambda}{\epsilon_0\hbar V_l} \label{eq:g_b}\\
\zeta_k=&-i\Xi \sqrt{\frac{\omega_k}{2\rho V_p v_k^2 \hbar}} \label{eq:zeta_k}
\end{align}

In Eqs. \ref{eq:g_a} and \ref{eq:g_b}, $\hat{\epsilon_\lambda}$ is the unit polarization vector of the $\lambda$ mode, $d_a=\bra{2}\mathbf{r}\ket{1}$ and $d_b=\bra{2}\mathbf{r}\ket{3}$ are the dipole moments with  $\bra{3}\mathbf{r}\ket{1}=0$. $V_l$ stands for the quantization volume for photons. However, we allow for electron phonon coupling between $\ket{1}$ and $\ket{3}$ and the coupling coefficient is defined by Eq. \ref{eq:zeta_k} \cite{imbusch_temperature_1964,Toyozawa2003} where $\Xi$ is the deformation potential, $\rho$ is the density $v_k$ is the group velocity of mode $k$, $V_p$ stands for the quantization volume for phonons. 

Using the equation of motion for a single time operator given by $\dot O(t) = (i\hbar)^{-1} [O,H]$, and using the Hamiltonian in Eqs. 1a-1d, we find the atom-field system evolves according to the following equation 

\begin{subequations}
\begin{align}
\dot{\sa{11}} =& -i \suml \glac \cld \sa{12} +i \sa{21} \suml \gla  \cl - i \sumk \zkc \bks \sa{13} +i \sumk \zk \sa{31} \bks \label{eq:eom_s11}\\
\dot{\sa{33}} =& -i \suml \glbc \cld \sa{32}+i \sa{23} \suml \glb \cl  -i \sa{31} \sumk \zk \bks + i\sumk \zkc \bks \sa{13} \label{eq:eom_s33}\\
\dot{\sa{21}} =& i\sa{21}\Omega_a- i\suml \glac \cld \left( \sa{22}-\sa{11} \right) +i  \suml \glbc \cld \sa{31} - i\sumk \zkc \bks \sa{23} \label{eq:eom_s21}\\ 
\dot{\sa{32}} =& -i\sa{32}\Omega_b-i \sa{31} \suml \gla \cl -i \left( \sa{33}- \sa{22} \right) \suml \glb \cl  +i \sumk \zkc \bks \sa{12} \label{eq:eom_s32}\\
\dot{\sa{31}} =& i\sa{31} (\Omega_a-\Omega_b) +i \suml \glac \cld \sa{32} +i \sa{21} \suml \glb \cl  -i \sumk \zkc \bks \left( \sa{33} -\sa{11} \right) \label{eq:eom_s31}\\
\dot{\cl}=& -i\omega_{\lambda} \cl -i \glac \sa{12} -i \glbc \sa{32} \label{eq:eom_cl}\\
\dot{\bk} =& -i\omega_k \bk -i \zk \sa{31} - i \zkc \sa{13} \label{eq:eom_bk} 
\end{align} \label{eq:eom}
\end{subequations}
Eqs. \ref{eq:eom_cl} and \ref{eq:eom_bk} can be integrated from initial time $t=0$ to give
\begin{subequations}
\begin{align}
\cl(t)=& c(0) e^{-i\omega_\lambda t} - i \glac \int_0^t \sa{12}(t') e^{-i\omega_\lambda \left(t-t'\right)} dt' - i \glbc \int_0^t \sa{32}(t') e^{-i\omega_\lambda \left(t-t'\right)} dt' \label{eq:clt}\\
\bk(t)=& b(0) e^{-i\omega_k t} - i \zk \int_0^t \sa{31}(t') e^{-i\omega_k \left(t-t'\right)} dt' - i \zkc \int_0^t \sa{13}(t') e^{-i\omega_k \left(t-t'\right)} dt' \label{eq:bkt}
\end{align}
\end{subequations}

Eqs. \ref{eq:clt} and \ref{eq:bkt} can be substituted into Eqs. \ref{eq:eom_s11}-\ref{eq:eom_s31} to express electronic density matrix operators as initial conditions of field operators. We first make the harmonic approximation to Eqs. \ref{eq:clt} and \ref{eq:bkt} to obtain 
\begin{subequations}
\begin{align}
\sa{12}(t')&=\sa{12}(t) e^{i\Omega_a(t-t')}\\
\sa{23}(t')&=\sa{23}(t)e^{-i\Omega_b(t-t')}\\
\sa{31}(t')&=\sa{31}(t) e^{-i(\Omega_a-\Omega_b)(t-t')}=\sa{31}(t) e^{-i\domega(t-t')} 
\end{align}
\label{eq:harmonic_approx}
\end{subequations}
The validity of this approximation is discussed in Whitley and Stroud [49] where we also assume that the time intervals are much shorter than the Rabi or natural lifetime for both the photon and phonon field.

Using the harmonic approximation in Eqs. \ref{eq:harmonic_approx}, we simplify the photon field in Eq. \ref{eq:clt} to 
\begin{equation}
\cl(t)	\simeq c(0) e^{-i\omega_\lambda t} \sa{12}(t) \int_0^t  e^{-i(\omega_\lambda-\Omega_a) \left(t-t'\right)} dt' - i \glbc \sa{32}(t) \int_0^t  e^{-i(\omega_\lambda-\Omega_b) \left(t-t'\right)} dt' \label{eq:clt2}
\end{equation}

since only the values of $t'$ within a few optical periods of $t$ are important to the integral.

Likewise, we apply the harmonic approximation to the phonon operator in Eq. \ref{eq:eom_bk} to obtain

\begin{equation}
\bk(t)	\simeq b(0) e^{-i\omega_k t} - i \zk \sa{31}(t)\int_0^t  e^{-i(\omega_k+\domega) \left(t-t'\right)} dt'  - i \zkc \sa{13}(t)\int_0^t  e^{-i(\omega_k-\domega) \left(t-t'\right)} dt' \label{eq:bkt2}
\end{equation}

As we did not adopt rotating wave approximation for the phonon field, the combined operator $\bks$ is more useful. We use Eq. \ref{eq:bkt2} to obtain
\begin{align}
\bk(t)+\bkd(t)=&\bk(0) e^{-i\omega_k t} + \bkd(0) e^{i\omega_k t}  + i\zk \sa{31} \int_0^t  e^{i(\omega_k-\domega) \left(t-t'\right)} -e^{-i(\omega_k+\domega) \left(t-t'\right)} dt' \nonumber\\ &-i \zkc \sa{13}\int_0^t  e^{-i(\omega_k-\domega) \left(t-t'\right)} -e^{i(\omega_k+\domega) \left(t-t'\right)}dt' \nonumber\\
\label{eq:bks}
\end{align}

assuming $\Delta\Omega=\Omega_a-\Omega_b$, $B(t)=\bk(0) e^{-i\omega_k t} + \bkd(0) e^{i\omega_k t}$ and $\Phi=\int_0^t e^{i(\omega_k-\domega) \left(t-t'\right)} - e^{-i(\omega_k+\domega) \left(t-t'\right)} dt' $. Note that Eq. \ref{eq:bks} is invariant with respect to its conjugate. We can thus ignore conjugation considerations on the phonon operators later in our derivation.

Substituting Eqs. \ref{eq:clt2} and \ref{eq:bks} into Eq. \ref{eq:eom_s11}, we obtain

\begin{align}
\dot{\sa{11}}=&i\sa{21} \suml \gla \cl(0) e^{-i\olm t} -i \suml \glac \cld(0) e^{i\olm t} \sa{12} +\lvert \gla \rvert^2 \sa{22} \int_{-t}^{t} e^{i(\olm-\Omega_a)(t-t')} dt' \nonumber\\ &+i \sa{31} \sumk \zk B(t) - i\sumk \zkc B(t) \sa{13} + \sumk \lvert \zk \rvert^2 \sa{33} (\Phi^*(t)+\Phi(t)) \nonumber\\
\simeq &i\sa{21} \suml \gla \cl(0) e^{-i\olm t} -i \suml \glac \cld(0) e^{i\olm t} \sa{12} + \suml \lvert \gla \rvert^2 \sa{22} 2\pi \dfn{\olm}{\oa} \nonumber\\&+ i \sa{31} \sumk \zk \bt - i\sumk \zkc \bt \sa{13} + \sumk \lvert \zk \rvert^2 \sa{33} 2\pi \left( \dfn{\ok}{\domega}-\dfnc{\ok}{\domega} \right) \nonumber\\
= & 2\Gamma_a \sa{22} + i\sa{21} \suml \gla \cl(0) e^{-i\olm t} -i \suml \glac \cld(0) e^{i\olm t} \sa{12} + 2\Gamma_p \sa{33}  + i \sa{31} \sumk \zk \bt - i\sumk \zkc \bt \sa{13} \label{eq:aom_s11_1}
\end{align}

where we have defined the damping constants for the $\ket{2}-\ket{1}$ transition to be 

\begin{equation}
\Gamma_a=  \suml \lvert \gla \rvert^2 \pi \dfn{\olm}{\oa} \label{eq:ga}
\end{equation}

 and the phonon damping to be 
 
 \begin{equation}
  \Gamma_p=  \sumk \lvert \zk \rvert^2 \pi \left( \dfn{\ok}{\domega}-\dfnc{\ok}{\domega} \right) \label{eq:gp}
 \end{equation}

We carry out the same procedure for Eqs. \ref{eq:eom_s11}-\ref{eq:eom_s31} to obtain

\begin{subequations}
\begin{align}
\dot{\sa{11}}= & 2\ga \sa{22} + i\sa{21} \suml \gla \cl(0) e^{-i\olm t} -i \suml \glac \cld(0) e^{i\olm t} \sa{12} + 2\gp \sa{33}  + i \sa{31} \sumk \zk \bt - i\sumk \zkc \bt \sa{13} \label{eq:aom_s11} \\
\dot{\sa{33}}=& 2\gb \sa{22} + i\sa{23} \suml \glb \cl(0) e^{-i\olm t} - i \suml \glbc \cld(0) e^{i\olm t} \sa{32} -  2\gp \sa{33} - i \sa{31} \sumk \zk \bt + i\sumk \zkc \bt \sa{13} \label{eq:aom_s33}\\
\dot{\sa{21}}=& \left( i \oa-\gb -\ga  \right) \sa{21} -i\suml \glac \cld(0)  e^{i\olm t} (\sa{22}-\sa{11}) + i \suml \glbc \cld(0) e^{i\olm t} \sa{31}  - i \sumk \zkc \bt \sa{23}  \label{eq:aom_s21}\\
\dot{\sa{32}}=& \left( -i \ob-\gb -\ga -\gp  \right) \sa{32} -i (\sa{33}-\sa{22}) \suml \glb \cl(0)  e^{-i\olm t} - i \sa{31} \suml \gla \cl(0) e^{-i\olm t}  + i \sumk \zkc \bt \sa{12}   \label{eq:aom_s32}\\
\dot{\sa{31}}=& \left( i \domega  -\gp \right) \sa{31} - i \suml \glac \cld(0)  e^{i\olm t} \sa{32} +\Gamma_{ab}\sa{22} \nonumber\\ &+ i \sa{21} \suml \glb \cl(0) e^{-i\olm t}  - i \sumk \zkc \bt (\sa{33}-\sa{11})-\gp' \sa{13}  \label{eq:aom_s31}
\end{align}
\label{eq:aom}
\end{subequations}

where  $\gb=  \suml \lvert \glb \rvert^2 \pi \dfn{\olm}{\ob}$, $\Gamma_{ab}=\pi \suml \glac \glb \left(  \dfn{\olm}{\ob} + \dfn{\olm}{\oa} \right)$ and $\gp'=\sumk {\zkc}^2 \pi \left( \dfn{\ok}{\domega}-\dfnc{\ok}{\domega} \right)  $. If we assume orthogonality of $\mathbf{d_a}$ and $\mathbf{d_b}$, $\Gamma_{ab}$ will vanish.

Now, we take the expectation value in a product of a monochromatic coherent state [49] by transforming

\begin{align*}
\chi_{ii}(t)&=\sa{ii}(t)\\
\chi_{31}(t)&=\sa{31}(t) e^{i(\oa-ob)t}\\
\chi_{32}(t)&=\sa{32} e^{-i\ob t}
\end{align*} 

such that Eq. \ref{eq:aom} become

\begin{subequations}
\begin{align}
\ave{\dot{\chi_{11}}}=& 2\ga \ci{22} + i\ci{21} G_a -i G_a^* \ci{12} + 2\gp \ci{33}  + i \ci{31} W - i W^* \ci{13} \label{eq:ave_s11} \\
\ave{\dot{\chi_{33}}}=& 2\gb \ci{22} + i\ci{23} G_b - i G_b^* \ci{32} -  2\gp \ci{33} - i \ci{31} W + i W \ci{13} \label{eq:ave_s33}\\
\ave{\dot{\chi_{21}}}=& \left( i \da-\gb -\ga  \right) \ci{21} -i G_a^* (\ci{22}-\ci{11}) + i G_b^* \ci{31}  - i W^* \ci{23}  \label{eq:ave_s21}\\
\ave{\dot{\chi_{32}}}=& \left( -i \db-\gb -\ga -\gp  \right) \ci{32} -i (\ci{33}-\ci{22}) G_b - i \ci{31} G_a + i W^* \ci{12}  \label{eq:ave_s32}\\
\ave{\dot{\chi_{31}}}=& \left( i (\da-\db)  -\gp \right) \ci{31} - i G_a^* \ci{32} + i \ci{21} G_b  - i W^* (\ci{33}-\ci{11})-\gp' \ci{13}  \label{eq:ave_s31}
\end{align}
\label{eq:ave}
\end{subequations}

where

\begin{subequations}
\begin{align}
G_a=& g_a^{\alpha} \ave{c_{\alpha}(0)} \label{eq:Ga}\\
G_b=& g_b^{\beta} \ave{c_\beta(0)}  \label{eq:Gb}\\
W=& \zeta_{\gamma} \ave{B_{\gamma}(0)} \label{eq:W}
\end{align}
\label{eq:GW}
\end{subequations}

. Furthermore, we assume a closed system such that 
\begin{equation}
\ci{11}+\ci{22}+\ci{33}=1 
\label{eq:sum}
\end{equation}

With Eq. \ref{eq:ave}, we can complete the set of system of equations in a $9\times9$ matrix $A$ where 
\begin{equation}
\frac{d}{dt}\mathbf{\Psi}(t)=A \mathbf{\Psi}(t) \label{eq:mat_eom}
\end{equation}
where $\mathbf{\Psi}(t)$ is a $9\times1$ column vector with $\Psi_1=\ci{33}$, $\Psi_2=\ci{32}$, $\Psi_3=\ci{31}$, $\Psi_4=\ci{21}$, $\Psi_5=\ci{11}$, $\Psi_6=\ci{12}$, $\Psi_7=\ci{13}$, $\Psi_8=\ci{23}$ and $\Psi_9=\ci{22}$. With this choice, the matrix $A$ becomes

\scriptsize
\begin{equation}
\begin{pmatrix}
-2\gp 	&	 -iG_a 	    &	 -iW 			&	0		&	0 		&	0		&	i W		&	i G_b 		&	2\gb		\\
-i G_b 	&-i\da-\ga-\gb-\gp &	 -iG_a 		&	0		&	0 		&	iW		&	0		&	0 			&	iG_b		\\
-i W 		& -iG_a 		    &	 i(\da-\db)-\gp 	&	iG_b		&	iW 		&	0		&	-\gp		&	0 			&	0		\\
0 		& 0	 		    &	 iG_b 	 	&i\da-\ga-\gb	&	iG_a 		&	0		&	0		&	-iW 			&	-iG_a		\\
2\gp 		& 0	 		    &	 iW	 	 	&iG_a 		&	0 		&	-iG_a		&	-iW		&	0 			&	2\ga		\\
0 		& iW	 		    &	 0	 	 	&  0	 		&	-iG_a		&-i\da-\ga-\gb	&	-iG_b		&	0 			&	iG_a		\\
iW 		& 0	 		    &	 -\gp	 	 	&  0	 		&	-iW		&-iG_b 		&-i(\da-\db)-\gp	&	iG_a 			&	0		\\
iG_b 		& 0	 		    &	 0	 	 	&  -iW	 	&	0		&	0		&	iG_a		&i\da-\ga-\gb-\gp 	&	-iG_b		\\
0 		& iG_b	 	    &	 0	 	 	&  -iG_a	 	&	0		&	iG_a		&	0		&	-iG_b 		&	-2(\ga+\gb)		
\end{pmatrix}
\label{eq:amat}
\end{equation}
\normalsize

where we have assumed that all coupling constants are real such that $G_a^*=G_a$, $G_b^*=G_b$, $W^*=W$ and $\gp'=\gp$. 

\section{Relationship between turning points and detuning}
Solving the steady state solution to Eq. \ref{eq:mat_eom} for $\ci{11},\ci{22},\ci{33}$, one obtains the population in each level $\rho_{11}$, $\rho_{22}$ and $\rho_{33}$. Here, we examine the coherent population trapping (CPT) case where $G_a=G_b=G$ and that $\ga=\gb=\Gamma_0$ which is the scaling factor for the entire system [49]. Thus, all units in our calculations are subjected to scaling factor $\Gamma_0$. In the CPT calculations, we assume that $\db=0$ and $\da$ is varied as described in the main text. In CPT experiments, $\rho_{22}$ is usually the main indicator in experiments and we also focus on this observable here. 
The solution for the excited state population $\rho_{22}$ under the condition $W=0,\gp=0$ is given in Eq. 2b in the main text. Here, we show the solution under which $\gp=0$ which gives 
\begin{equation} 
\rho_{22}(\db,W)=\frac{2\db^2 G^2}{\db^4+2\db^3 W +8 \db W (W^2-G^2) +2 \db^2 (4+G^2+3W^2)+8(G^4-2(G^2-2)W^2+W^4)} \label{eq:r22_gp0}
\end{equation}
The turning points for Eq. \ref{eq:r22_gp0} can be obtained by taking the equating first order derivative to zero. One of the roots will be at $\da=0$ which is the same as CPT. Two roots are imaginary and two roots are real, one for the turning point for $\da>0$ and one for $\da<0$. Here, we Taylor expand the roots for small $W$ to the fourth power to obtain
 \begin{align}
\delta_{b,max}\simeq & \pm2^{3/4} G - \frac{1}{2} \left( \frac{1}{2}+\frac{1}{\sqrt{2}} \right) W \pm \frac{2^{1/4}(64\sqrt{2}+G^2-30\sqrt{2}G^2)}{64G^3}W^2+\frac{64+G^2}{64\sqrt{2}G^4}W^3 \nonumber\\
& \mp \frac{2^{1/4}((2037\sqrt{2}-152)G^4+(6656\sqrt{2}-1792)G^2-40960\sqrt{2})}{16384G^7}W^4  \label{eq:r22_series}
\end{align}
where when $W\rightarrow0$, we obtain the turning points at $\pm\pm2^{3/4} G$ as described in Fig. 1(c) of the main text. The linear shift term in Eq. \ref{eq:r22_series} in $- \frac{1}{2} \left( \frac{1}{2}+\frac{1}{\sqrt{2}} \right) W$ is independent of the sign of the root and is valid for small $W$ as shown in Fig. 2.

\section{Time dependence of population} 
\label{sec:time}
The time dependence of the population is obtained by solving the eigenvalues $s_m$, eigenvectors $\mathbf{\nu}(m)$ of matrix $A$ in Eq. \ref{eq:amat} and its reciprocal eigenvector $\mathbf{w}(m)$. The time dependent solution of Eq. \ref{eq:mat_eom} is then given by [49]

\begin{equation}
\Psi_i(t)=\sum_{m=1}^9 \sum_{j=1}^9 \nu_i(m) w_j(m) \Psi_j(0) e^{s_m t} \label{eq:time}
\end{equation}  
 where $\mathbf{\Psi}(0)$ is a $9\times1$ vector of the initial condition. Here, we solved for $\Psi_1(t)=\rho_{33}(t)$, $\Psi_5(t)=\rho_{11}(t)$ and $\Psi_9(t)=\rho_{22}(t)$ as shown in Fig. 3 under the initial condition that $\Psi_5(0)=1$ and all other terms are zero i.e. all population are in the ground state at $t=0$.

\bibliographystyle{alpha}